\documentclass[runningheads]{llncs}

\usepackage{graphicx}
\usepackage{amsmath,amssymb,amsfonts}
\usepackage{cite}
\usepackage{hyperref}
\usepackage{url}
\usepackage{booktabs}
\usepackage{multirow}
\usepackage{array}
\usepackage{float}
\usepackage{caption}
\usepackage[table]{xcolor}
\usepackage{tikz}
\usepackage{subcaption}
\usepackage{xcolor}
\usepackage[table]{xcolor}
\usepackage{colortbl}
\usepackage{array}
\usepackage{longtable}
\usepackage{booktabs}
\usepackage{array}
\usepackage{qcircuit}
\usepackage{tabularx}
\usepackage{fullpage}

\definecolor{EScolor}{RGB}{46,125,50}     % Dark green
\definecolor{FUcolor}{RGB}{129,199,132}   % Medium green
\definecolor{SUcolor}{RGB}{255,241,118}   % Light yellow
\definecolor{NUcolor}{RGB}{224,224,224}   % Light gray

\definecolor{EScolor}{HTML}{FFD700} % Gold color for ES
\newcommand{\ket}[1]{\left|#1\right\rangle}

\hypersetup{
    colorlinks=true,
    linkcolor=blue,
    citecolor=blue,
    urlcolor=blue
}
\newcommand{\ES}{\cellcolor{EScolor}\textbf{ES}}
\newcommand{\FU}{\cellcolor{FUcolor}FU}
\newcommand{\SU}{\cellcolor{SUcolor}SU}
\newcommand{\NU}{\cellcolor{NUcolor}NU}

\begin{document}

\title{Quantum Software Architecture Framework (QSAF): A Component-Based Framework for Designing Hybrid Quantum–Classical Systems}

\author{
Arvind W. Kiwelekar\inst{1} %\thanks{Corresponding author} \and
Shweta Tembe\inst{1} \and
Uzma G. A. Munde\inst{1} \and
Siddhesh Jadhav\inst{1} \and
Manjushree D. Laddha\inst{1} \and
Harsha R. Gaikwad\inst{1}
}

\institute{
$^{1}$Center for Excellence in Quantum Technology \\ 
Dr. Babasaheb Ambedkar Technological University %\\
Lonere 402103, Maharashtra, India \\
\email{awk@dbatu.ac.in}
}

\maketitle

\begin{abstract}Quantum software development has largely focused on algorithms, with limited attention to software architecture. As computing moves toward hybrid quantum–classical systems, this gap limits scalability, reusability, and engineering rigor.

This study introduces a component-based quantum software architecture framework (QSAF) for hybrid quantum-classical software systems, enabling developers to transition from circuit-level design to system-level reasoning. We identified 34 reusable quantum circuit primitives across seven functional categories and reinterpreted them as architectural components with explicit interfaces and design-relevant constraints. These components are further characterized using nonfunctional dimensions, such as circuit depth, error sensitivity, and information flow, enabling a structured analysis of design trade-offs.

The proposed QSAF framework establishes a multilevel abstraction hierarchy linking quantum gates, circuit primitives, algorithmic structures, and hybrid system architectures. Through this approach, common workflows,  particularly hybrid quantum–classical workflows, such as variational quantum algorithms, can be systematically decomposed, compared, and optimized.

By making the architectural structure and trade-offs explicit, this study provides a foundation for quantum software engineering, supporting modular design, reuse, and informed architectural decision-making in quantum-application development.
\end{abstract}


\section{Introduction}
Quantum computing is rapidly evolving from algorithmic demonstrations to software-intensive systems that integrate quantum and classical components into increasingly complex workflows. However, current quantum software development remains largely circuit-centric, with most tools, languages, and methodologies focusing on constructing and optimizing individual quantum circuits rather than supporting higher-level system design. Consequently, key software engineering concerns, such as modularity, composability, reuse, and architectural reasoning, have not been systematically addressed. This gap is particularly significant in hybrid quantum–classical systems, where multiple interacting components, feedback loops, and hardware constraints must be coordinated. The absence of architectural abstractions limits developers' ability to reason about the system's structure, evaluate design alternatives, and manage nonfunctional trade-offs.

This study addresses this gap by adopting  an intentional architectural approach   to quantum systems. Intentional architecture  ~\cite{bass2012software} has long been recognized as a guiding principle in classical software engineering \cite{pantoja2024training, kiwelekar2014ontological}. Explicit architectural structuring supports the maintainability and reusability of the software. 
 
 Furthermore, this study argues that achieving an intentional quantum software architecture requires reframing  reusable circuit constructs as architectural components. To this end, we surveyed representative algorithm families \cite{shor1994algorithms}, \cite{peruzzo2014variational}, \cite{farhi2014quantum}, \cite{1629135}, \cite{khan2025quantum},\cite{10646500} and performed algorithmic decomposition to identify modular and composable circuit structures that recur across applications. These structures are considered reusable architectural primitives. Building on this analysis, we propose a multidimensional classification framework inspired by the software connector taxonomy~\cite{mehta2000towardsss}. The framework characterizes quantum circuit components along both functional and nonfunctional dimensions. 

 Unlike prior studies that primarily focus on algorithm design and   implementation tooling, this study does not introduce new quantum algorithms or circuit constructions. Instead, it focuses on the architectural organization of recurring circuit structures that appear across multiple algorithmic families. By treating these structures as first-class architectural components with explicit interfaces and quality-relevant properties, this study shifts the focus from algorithmic functionality to the architectural structuring of quantum software systems.

 This study makes the following contributions.
\begin{enumerate}
    \item \textit{Architectural Reframing of Quantum Circuits:}
We conceptualize recurring quantum circuit structures as architectural components with defined interfaces, facilitating reasoning beyond algorithm-centric representation.

\item \textit{Component Taxonomy of Quantum Circuit Primitives:}
We identified and organized 34 reusable circuit primitives into seven functional categories, thereby providing a structured vocabulary for describing quantum software elements.

\item \textit{multidimensional Characterization Framework:}
We establish a set of orthogonal nonfunctional dimensions (e.g., complexity, information flow, and error sensitivity) that aid in reasoning about the design trade-offs in quantum systems.

\item \textit{Abstraction Hierarchy for Quantum Software Systems:}
We propose a multilevel abstraction model that connects gate-level constructs to higher-level architectural compositions, thereby enabling system-level reasoning.

\item \textit{Foundation for Architecture-Oriented Design Exploration:}
This study lays a conceptual foundation for future research exploring how architectural abstractions can support modularity, reuse, and design decision-making in quantum software systems, rather than providing an empirical validation.

\end{enumerate}

\begin{figure}[t]
\centering
\includegraphics[width=0.9\textwidth]{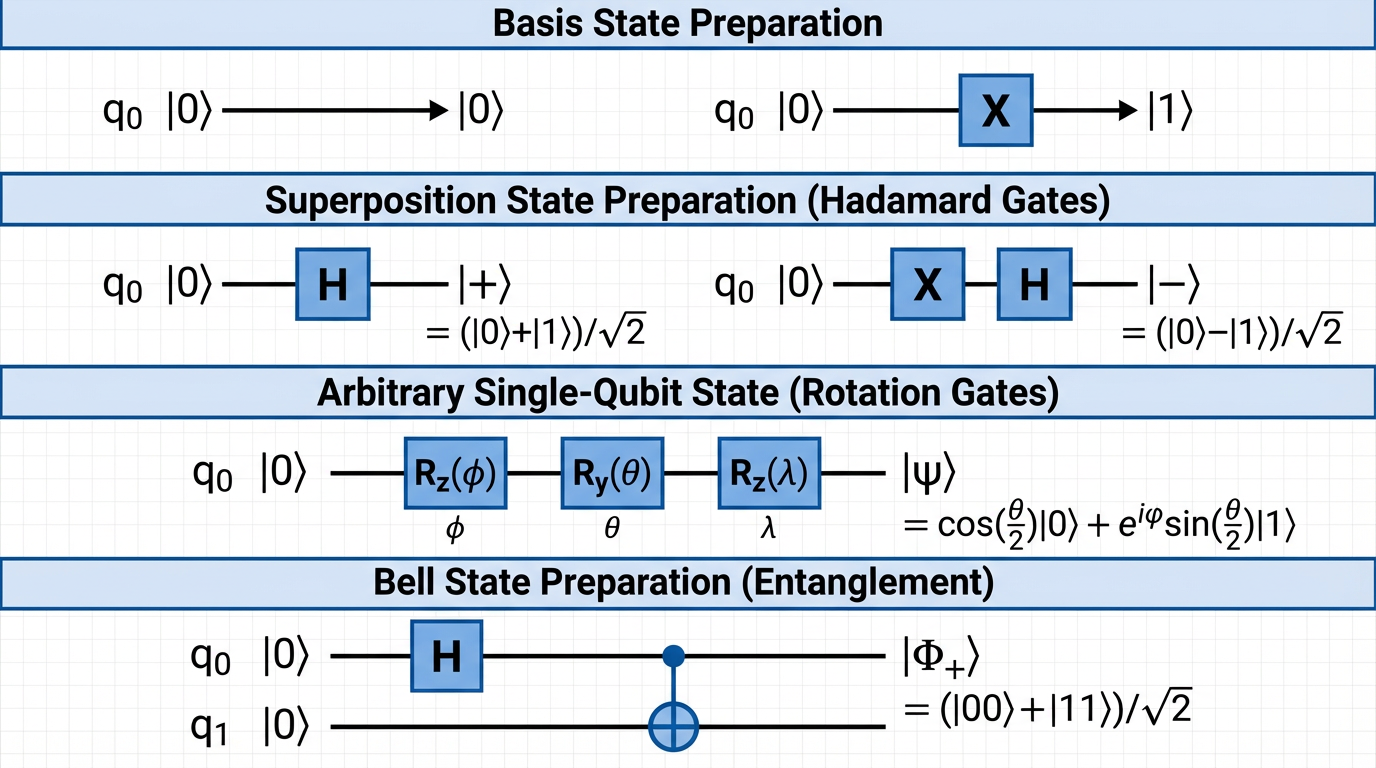}

\centering{ $H$: Hadamard Gate, $\oplus :$ CNOT   gate, $R_{*}: Rotational Gate$, $X$ Pauli X gate}
\caption{Quantum circuit diagrams for state preparation}
\label{fig:state_prep}
\end{figure}

\section{Background and Earlier Work}
The work presented in this study is at the intersection of quantum algorithms and software architectures. Quantum circuits are the fundamental units of quantum software systems\cite{gill2022quantum}. They are built around qubits and quantum gates. Unlike classical logic circuits, quantum circuits manipulate \textit{qubits} through unitary transformations, such as the Pauli X (NOT) gate, rotational gates, and controlled NOT gates.  A qubit may be in the state of $\ket{0}$, $\ket{1}$, or a superposition state, that is, $\alpha\ket{0} + \beta\ket{1}$.  A few examples of \textit{quantum circuits } are shown in Figure \ref{fig:state_prep}.  They use the Hadamard, C-NOT, Pauli X, $R_{x}$, $R_{y}$, and $R_{z}$ gates  (rotations around the x, y, and z axes) to create various types of quantum states. 
\textit{Quantum algorithms} employ quantum-mechanical phenomena, such as superposition, entanglement, and interference, to solve problems~\cite{nielsen2010quantum}. The gate-based paradigm of quantum computing uses discrete quantum gates that are applied to qubits.  The solution to a problem, such as searching, factorization, or optimization, is realized as a combination of quantum gates, called a quantum circuit or algorithm. For example, Grover's algorithm ~\cite{grover1996fast} achieves a quadratic speedup for an unstructured search.

%Software architecture captures the high-level structure of software systems in terms of components, their relationships, and principles governing their design~\cite{bass2021software}. Numerous architectural principles guide the architectural design. However, this study focused on five architectural principles. (i) \textit{Modularity and Separation of Concerns}: Decomposing systems into cohesive modules with well-defined interfaces. (ii) \textit{Abstraction and Information Hiding}: Implementation details are hidden behind abstract interfaces. (iii) \textit{Composition and Composability}: Complex systems are built by composing simpler components according to well-defined rules. (iv) \textit{Reusability}: Design components for use across multiple contexts.

Earlier approaches that address software engineering concerns are reviewed in the following sections.  
\begin{enumerate}
    \item  \textbf{Classification and Cataloguing:}
A study  \cite{10646500} identified patterns in variational ansatzes by focusing on a single algorithm family.  This study analyzed the reuse strategies across  quantum-classical optimization algorithms.   A modular quantum library \cite{10.1145/2629430} has been proposed,   
which generates quantum circuits for the well-known quantum algorithms. This study mainly addressed low-level implementation concerns.
Reference \cite{arnault2024} classified approximately 130 quantum algorithms according to the fundamental mathematical problems they solved, their real-world applications, and the main subroutines they employed. In \cite{martyn2021grand}, the authors introduced the unification of frequently used quantum subroutines, such as amplitude amplification, phase estimation, and amplitude estimation, via quantum singular value transformation (QSVT).
\item  \textbf{Reusability:} Reusability is another high-level concern that is addressed at the algorithmic level.  These approaches devise novel techniques to advance reuse at the qubit level \cite{PhysRevX.13.041057}, promote reusable templates for compiling quantum circuits
\cite{avitabile2022, 10.1145/3670417},  adopt reusable optimized code generation strategies \cite{9774510},   support reusable components across algorithm families \cite{Carrazza_2023}, develop modules for specific purposes such as for the variational quantum eigensolver \cite{Ghasempouri2023},  support quantum ancilla reuse for modular quantum programs \cite{9138979},  and  leverage parameterized circuit execution \cite{rajagopala2025}. These approaches aim to reduce hardware resource requirements \cite{PhysRevX.13.041057} and   qubit counts \cite{hua2023exploitingqubitreusemidcircuit}, and emphasize extensibility and abstraction \cite{Carrazza_2023}.
\item  \textbf{Separation of Concerns:} Designing software systems by separating  concerns and encapsulating them in a separate module is a guiding principle in classical software engineering \cite{thakare2016skiplpa}. Quantum software systems also follow this principle.  
A platform-agnostic modular architecture for quantum benchmarking was proposed in \cite{patel2025platformagnosticmodulararchitecturequantum}. It proposes a modular benchmarking architecture that is decoupled from the specific platforms.   
An efficient parameterized compilation technique for hybrid quantum programming was presented in \cite{10313890}.
This enhances the architectural separation between the structure and parameters. 
\item  \textbf{Generative approaches}: Software construction methods intend to automate the software development process.
A model-driven framework for composition-based quantum circuit design was presented in \cite{10.1145/3688856}. The work reported in
\cite{10313874}  describes the synthesis of approximate parametric circuits for variational quantum algorithms. Reference  \cite{9651462} introduced an open-source framework for optimal circuit synthesis. This tool integrates formal methods into the circuit construction.  Reference \cite{10.1145/3589014.3595066} provides an overview of Intel’s quantum software stack. The SDK supports modular circuit construction and execution.  A modular component-based quantum circuit synthesis approach was presented in \cite{10.1145/3586039}. This approach decomposes the circuits into reusable building blocks. 
\end{enumerate}

\subsection{Architectural Problem}

Although quantum algorithms have been extensively studied, the architectural
structure of quantum software systems is relatively unexplored.
Most current implementations are expressed directly as circuit-level
descriptions of specific algorithms or hardware toolchains.
Consequently, reusable circuit constructs are rarely treated as
architectural components with explicit interfaces, well-defined
responsibilities, and quality attributes relevant to system design.

From a software architecture perspective, this raises a fundamental
question: how can quantum software systems be structured using
reusable architectural elements rather than an ad hoc circuit composition?
Addressing this question requires identifying recurring circuit
structures, abstracting them as architectural components, and
characterizing their compositional roles and quality-relevant
properties. Therefore, in this study we develop an
architecture-oriented abstraction framework that enables architects
to  systematically reason about modularity, reuse, and design
trade-offs in quantum software systems.

\begin{table}[h!]
\centering
\caption{Heatmap Representation of Primitive Usage Across Algorithms}
\label{tab:heatmap}

\begin{tabular}{llccccc}
\hline
\textbf{No.} & \textbf{Primitive} & \textbf{Grover} & \textbf{Shor} & \textbf{VQE} & \textbf{QAOA} & \textbf{Sim} \\
\hline

\multicolumn{7}{l}{\textit{State Preparation}} \\
1 & Basis States & \ES & \ES & \ES & \ES & \ES \\
2 & Superposition (H) & \ES & \ES & \ES & \ES & \ES \\
3 & Arbitrary States & \ES & \ES & \ES & \ES & \ES \\
4 & Bell States & \ES & \ES & \ES & \ES & \ES \\
5 & GHZ States & \SU & \SU & \SU & \SU & \ES \\
6 & Cluster States & \NU & \NU & \NU & \NU & \ES \\

\multicolumn{7}{l}{\textit{Entanglement Generation}} \\
7 & Bell State Circuits & \FU & \FU & \FU & \FU & \FU \\
8 & GHZ State Circuits & \FU & \FU & \FU & \FU & \FU \\
9 & W State Circuits & \FU & \FU & \FU & \FU & \FU \\
10 & Cluster State Circuits & \NU & \NU & \NU & \NU & \FU \\

\multicolumn{7}{l}{\textit{Amplitude Amplification}} \\
11 & Grover Operator & \ES & \NU & \NU & \NU & \NU \\
12 & Diffusion Operator & \ES & \NU & \NU & \NU & \NU \\
13 & Reflection Operators & \ES & \NU & \NU & \NU & \NU \\
14 & Amplitude Damping & \NU & \NU & \NU & \NU & \NU \\

\multicolumn{7}{l}{\textit{Basis Transformation (QFT)}} \\
15 & Standard QFT & \NU & \ES & \NU & \NU & \SU \\
16 & Inverse QFT & \NU & \ES & \NU & \NU & \SU \\
17 & Approximate QFT & \NU & \ES & \NU & \NU & \SU \\

\multicolumn{7}{l}{\textit{Oracle Construction}} \\
18 & Phase Oracles & \ES & \ES & \NU & \NU & \NU \\
19 & Bit-Flip Oracles & \ES & \NU & \NU & \NU & \NU \\
20 & Arithmetic Oracles & \NU & \ES & \NU & \NU & \NU \\
21 & Boolean Oracles & \ES & \NU & \NU & \NU & \NU \\

\multicolumn{7}{l}{\textit{Phase Estimation}} \\
22 & Standard QPE & \NU & \ES & \NU & \NU & \SU \\
23 & Iterative QPE & \NU & \ES & \NU & \NU & \SU \\
24 & Bayesian QPE & \NU & \ES & \NU & \NU & \SU \\

\multicolumn{7}{l}{\textit{Variational Ansatz}} \\
25 & Hardware-Efficient Ansatz & \NU & \NU & \ES & \ES & \NU \\
26 & Problem-Inspired Ansatz & \NU & \NU & \ES & \ES & \NU \\
27 & UCCSD Ansatz & \NU & \NU & \ES & \NU & \NU \\
28 & Heuristic Ansatz & \NU & \NU & \ES & \ES & \NU \\
29 & Hamiltonian Ansatz & \NU & \NU & \ES & \NU & \ES \\

\multicolumn{7}{l}{\textit{Auxiliary Operations}} \\
30 & SWAP Gates & \SU & \SU & \SU & \SU & \SU \\
31 & Controlled Operations & \ES & \ES & \ES & \ES & \ES \\
32 & Toffoli Gates & \ES & \SU & \SU & \SU & \SU \\
33 & Measurement & \ES & \ES & \ES & \ES & \ES \\
34 & Ancilla Management & \SU & \SU & \SU & \SU & \SU \\

\hline
\end{tabular}

\noindent
\textbf{Legend:}
\colorbox{EScolor}{\textbf{ES}} Essential \quad
\colorbox{FUcolor}{FU} Frequently Used \quad
\colorbox{SUcolor}{SU} Sometimes Used \quad
\colorbox{NUcolor}{NU} Not Used
\end{table}

\section{Methodology}
A two-step method that integrates a  literature review and algorithmic decomposition analysis was used to identify reusable quantum circuits.  First, we performed a structured multi-database literature search using defined queries.
The databases included Google Scholar, SciSpace Full Text and arXiv. The search queries, namely, "\textit{quantum circuit primitives}", "\textit{quantum algorithm building blocks}", "\textit{quantum software engineering design patterns}", and "\textit{quantum circuit library module composition}",  were used to obtain relevant papers.
The inclusion criteria were defined to include only gate-based quantum computing, circuit-level primitives, target algorithm coverage,  and peer-reviewed publications. The exclusion criteria were applied to discard articles that focused solely on hardware, used non-gate-based paradigms, lacked sufficient technical details, or were duplicates. 

The selected publications covered four types of algorithms: searching, order finding, optimization, and simulation. In the second step, we analyzed complete circuit implementations to identify functional blocks with distinct purposes and elevated them to reusable components if they functioned independently and appeared in multiple algorithms.  Furthermore, each selected publication was evaluated for stakeholder needs, orthogonality, and practical utility to identify classification dimensions. This analysis resulted in nine orthogonal dimensions, seven functional categories, and 34 primitives. 

The proposed taxonomy was evaluated using a multi-stage validation procedure. (i)\textit{Expert agreement study.}
Six domain experts with experience in quantum computing and software engineering independently classified the 34 identified primitives into seven functional categories.  The participants were provided with concise descriptions of each primitive. No collaboration was permitted during classification.
Inter-rater reliability was assessed using Fleiss’ $\kappa$ for multi-rater agreement. The resulting agreement score was $\kappa = 0.86$, indicating substantial agreement.  
%Overall raw agreement exceeded 85\% across all primitives. 
Disagreements were concentrated primarily between the \emph{state preparation} and \emph{entanglement generation} categories, accounting for most of the classification variance. Minimal cross-category confusion was observed among other functional groups. (ii)\textit{Decision-rule consistency check.}
To verify mutual exclusivity, each primitive was evaluated against the ordered decision tree described in Section 4.1. For all primitives, exactly one criterion was satisfied, confirming the deterministic label assignment.   (iii)\textit{Threats to validity} The results of the study presented here may be affected by various factors, including construct, internal, and  external validity.  Although representative algorithms from search, periodicity, variational optimization, and simulation were examined, other emerging
quantum algorithm paradigms may contain additional reusable circuit
structures. While the participating experts had experience in quantum computing and software engineering, subjective judgement may
influence the assignment of primitives to categories.  As quantum computing
 evolves, new circuit abstractions may
emerge. 

%\paragraph{External Applicability Assessment.}
%To examine coverage beyond the algorithm families used during extraction, the taxonomy was applied to two additional representative circuits not used in the initial identification process: the Harrow–Hassidim–Lloyd (HHL) algorithm and a discrete-time quantum walk construction. In both cases, all the constituent circuit structures were successfully mapped to existing categories without requiring category extension or modification. This suggests that the taxonomy generalizes beyond the original algorithm’s sample.

\begin{table*}[t]
\centering
\caption{Interface Definitions Functional Categories}
\label{tab:module_interface}
\small
\resizebox{\textwidth}{!}{%
\begin{tabular}{p{2.cm}p{2.1cm}p{2.2cm}p{1.5cm}p{2.4cm}p{2cm}}
\hline
\textbf{Category} & $\mathbf{Q_{in}}$ & $\mathbf{Q_{out}}$ & $\mathbf{Q_{anc}}$ & $\mathbf{(U, P)}$ & \textbf{Information Flow} \\ \hline

State \newline  Preparation &
null or basis state &
Prepared Quantum state &
Optional &
Fixed unitary; structural  &
Local $\rightarrow$ Global \\ 

Entanglement Generation &
Independent qubits &
Correlated  qubits &
Optional &
unitary; structural &
Global  \\ 

Oracle \newline Construction &
Problem \newline register &
Marked states &
Often \newline required &
Problem-dependent  &
Encoded in problem  \\ 

Amplitude \newline Amplification &
Superposed state &
Amplitude-reshaped state &
Optional &
Reflection &
Global interference \\ 

Basis Transformation  &
Computational basis state &
Phase/ frequency basis state &
None &
Fourier unitary; fixed &
Global spectral transformation \\ 

Phase \newline Estimation &
Eigenstate and control register &
Measured phase register &
Required &
Controlled-$U$; fixed  parameters &
control $\rightarrow$ target \\ 

Variational Ansatz &
Parameterized quantum state &
Parameterized quantum state &
Optional &
Parameterized unitary; variational  &
quantum-classical loop \\ \hline

\end{tabular}}
\end{table*}

\section{Architectural Component Model for Quantum Software}
 
In this study, we conceptualized  a quantum circuit  as  a module with the  following  architectural interface definition and constraints:

\begin{equation}
M = (Q_{\text{in}}, Q_{\text{out}}, Q_{\text{anc}}, U, \mathcal{P})
\end{equation}

where $Q_{\text{in}}$ is the input qubits, $Q_{\text{out}}$ is the output qubits, $Q_{\text{anc}}$ is the ancilla qubits, $U$ is the unitary operator, and $\mathcal{P}$ is the parameter.  
$\mathcal{P}$ captures the structural parameters that determine the circuit topology, qubit allocation, and composition structure, and the variational parameters that correspond to runtime-tunable numerical values.
Unlike classical architectures, these modules must satisfy the constraints imposed by quantum phenomena. 
These are (1) \textit{No-Cloning Constraint}: modules cannot duplicate quantum data, requiring explicit state shaping. (2) \textit{Unitarity constraint}: All modules encapsulate  reversible transformations.  (3)
\textit{Entanglement constraint}: Modules cannot be fully isolated. Entanglement may cross boundaries; hence,  explicit entanglement contracts are required at the interfaces.  Although the abstract module schema remains uniform, each functional category 
instantiates this interface differently in terms of input/output structure, 
ancilla usage, parameterization, and information-flow characteristics, 
The summary is presented in Table~\ref{tab:module_interface}.
This schema defines a quantum component in terms of its interface (input/output qubits), internal resources (ancilla), transformation logic (unitary), and parameterization.
\begin{figure}[t]
\centering
\includegraphics[width=\textwidth]{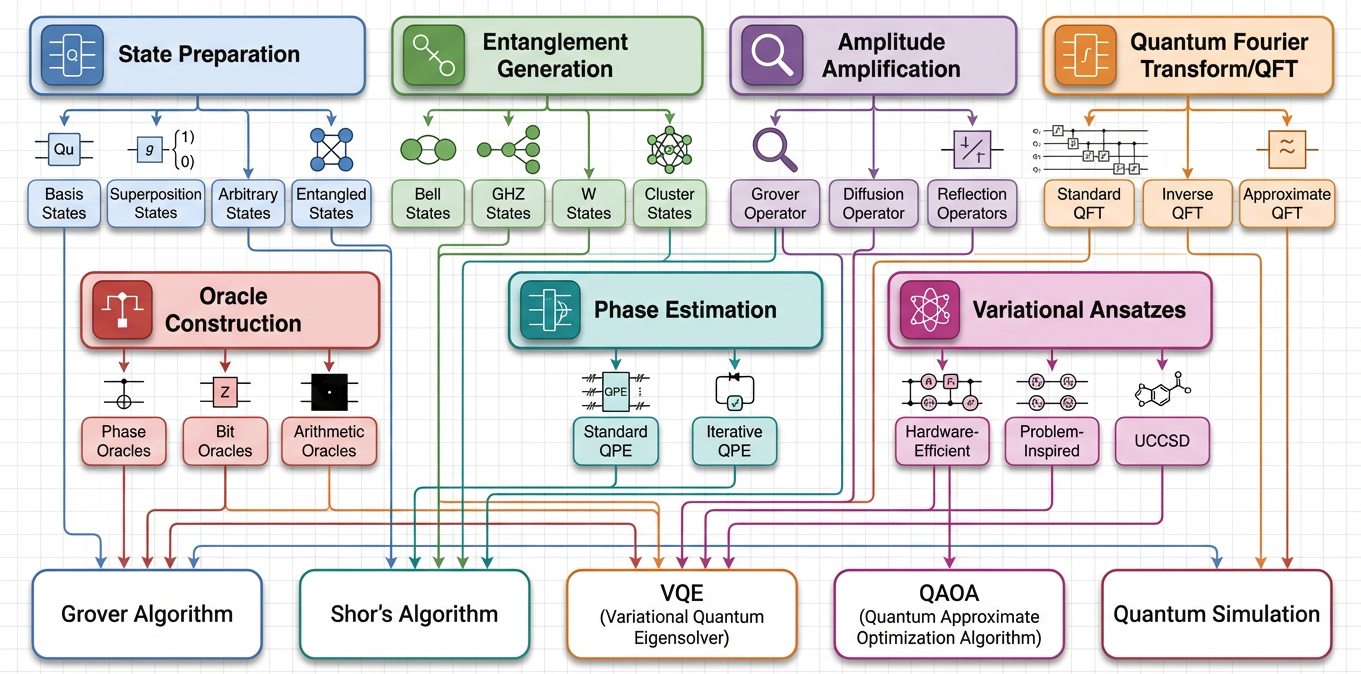}
\caption{Taxonomy of architectural components in quantum software systems, illustrating reusable circuit structures across algorithm families.}
\label{fig:taxonomy}
\end{figure}

\begin{table*}[t]
\centering
\caption{Functional Categories, Classical Analogies, and Circuit Complexity}
\label{tab:arch_categories}
\small
\resizebox{\textwidth}{!}{%
\begin{tabular}{p{2.8cm}p{3.3cm}p{3cm}p{3.2cm}}
\hline
\textbf{Functional \newline Category} & \textbf{Dominant Concern} & \textbf{Classical Analogy} & \textbf{Circuit Complexity} \\ \hline

State Preparation &
State preparation logic &
Object factory &
$O(1)$ to $O(n)$  \\ 

Entanglement \newline Generation &
Global interactions among the qubits. &
Coordination \newline mechanism &
$O(n)$ to $O(n+|E|)$  \\ 

Amplitude \newline Amplification &
Marking of good states &
Iterative refinement loop &
$O(\sqrt{N})$ iterations; $O(n)$ gates \\ 

Basis \newline Transformation  &
Basis conversion  to reveal global structure &
FFT &
$O(n^2)$, $O(n \log n)$  \\ 

Oracle \newline Construction &
Encode problem-specific predicates &
Predicate evaluator, Interfaces &
Polynomial in input size \\ 

Phase Estimation &
Extracting phase information &
Eigenvalue solver &
$O(n^2)$  \\ 

Variational Ansatz &
Parameterised circuit template for hybrid optimisation workflows &
Template-based optimisation model &
$O(n)$ to $O(n^2)$ \\ \hline

\end{tabular}}
\end{table*}
\subsection{Categories of Architectural Components}

\begin{figure}[t]
    \centering
    \includegraphics[scale=0.24]{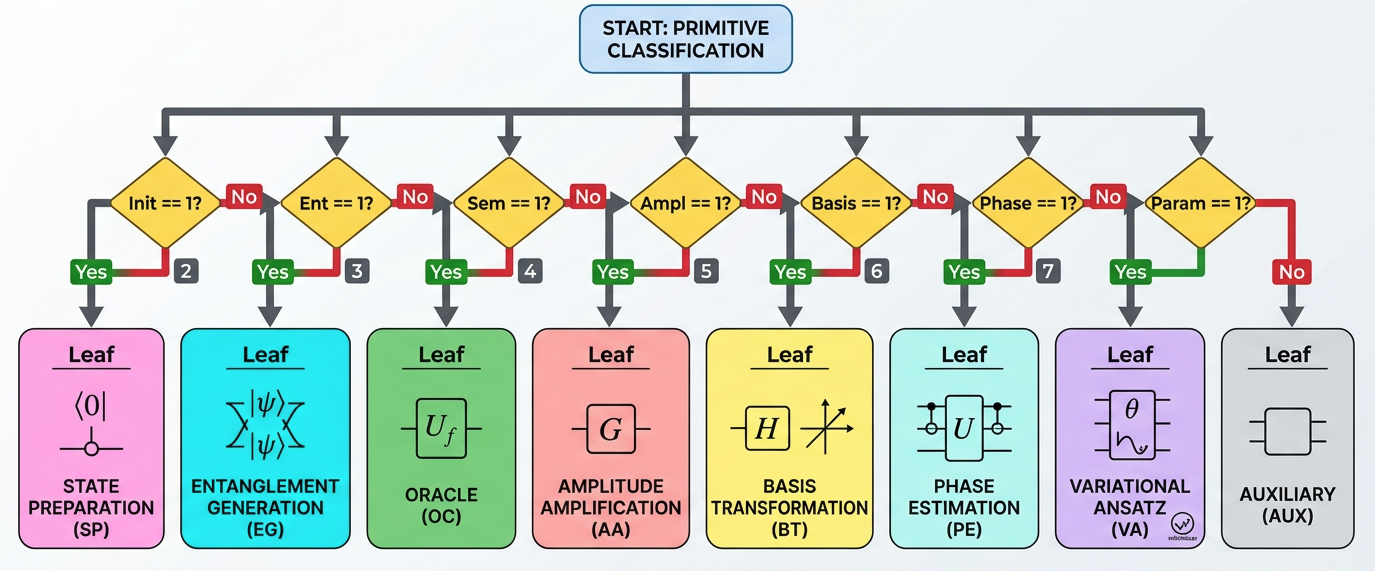}
    \caption{A Decision-tree for classifying quantum circuit primitives.}
    \label{class}
\end{figure}

Through systematic analysis, we identified 34 architectural components organized into seven functional categories, as shown in Figure~\ref{fig:taxonomy}, which can be viewed as a taxonomy.  
The taxonomy defines seven architectural element types that capture distinct functional responsibilities in gate-based quantum software systems: (i)\textit{State Preparation (SP)}, (ii)\textit{Entanglement Generation (EG)}, (iii)\textit{Oracle Construction (OC)}, (iv)\textit{Amplitude Amplification (AA)}, (v)\textit{Basis Transformation (BT)}, (vi)\textit{Phase Estimation (PE)}, and (vii)\textit{Variational Ansatz (VA)}. The quantum circuits included in the {\em auxiliary operations} from Table \ref{tab:heatmap} are straightforward and are realized as individual quantum gates. 
These categories are tabulated in Table \ref{tab:heatmap}  along with the classes of algorithms in which they are  used. The usage labels in Table \ref{tab:heatmap} were assigned based on the frequency with which a primitive appeared in the analyzed circuit implementations.  Furthermore, Table \ref{tab:arch_categories} compares and differentiates these categories  based on the dominant concern encapsulated by them,  a similar design abstraction from classical software systems, and circuit complexity measured using quantum gates. Their primary role is to establish the initial quantum state on which subsequent computations operate.  
\begin{enumerate}
    \item \textbf{State Preparation Components:}
 These circuits mediate algorithmic assumptions about the initial state conditions and concrete gate-level realizations. As illustrated in Figure \ref{fig:state_prep}, this class encompasses circuit structures, including basis state initialization, superposition state preparation, arbitrary single-qubit state construction via parameterized rotations, and multi-qubit entangled-state generation. 
Functionally, their role resembles that of factory-like abstractions from object-oriented-design patterns. However, unlike classical initialization or object factories, state preparation operates on a global, non-observable quantum state.
\item \textbf{Entanglement Generation Components:}
These components establish non-classical correlations among the qubits.  This category includes recurring structures, such as Bell, GHZ, W, and cluster-state generators, each producing characteristic entanglement patterns. Their implementations vary in depth and connectivity, ranging from $ O (1) $ to $ O (n +|E |) $.   However, their architectural roles remain unchanged.   Entanglement generators create nonlocal, nonseparable quantum correlations that fundamentally alter the global system state.

\item \textbf{Amplitude Amplification Components}
  These components encapsulate the logic for iteratively amplifying the target or good states while suppressing the non-target or bad states. The circuit complexity depends on the number of Grover iterations, which is typically $O(\sqrt{N})$, with each iteration composed of a polynomial-size quantum circuit.    This category includes recurring structures, such as Grover, diffusion, reflection, and amplitude estimation operators.   By isolating amplitude manipulation from problem encoding and algorithm control flow, these components promote a separation of concerns and allow architects to explicitly reason about the probabilistic amplification behavior within quantum software systems. 
  
  %Figure~\ref{fig:amplification} illustrates representative amplitude amplification circuit structures.

\item \textbf{Basis transformation  components  }
  These components transform assumptions expressed in one basis and their realization in the other. This reveals global properties, such as periodicity or phase relationships, that are not explicitly visible.  The basis transformation is realized independently of the problem-specific logic in terms of complex global interference patterns.  The circuit complexity ranges from  $O(n^2)$ gates to  $O(nlogn)$. This category includes the standard quantum Fourier transform (QFT), its inverse, and approximate variants, which recur across algorithms that rely on frequency or period extraction.  By isolating basis-transformation concerns from algorithm control and problem encoding, these components support architects in reasoning explicitly about representation changes in quantum software systems. Classically, basis transformation components are analogous to spectral transformations, such as the fast Fourier transform (FFT), which converts data from the time or spatial domain into the frequency domain.

\item \textbf{Oracle Construction}:  These components encode problem-specific information into a quantum computation. Rather than performing computations, these components embed decision predicates, constraints, and arithmetic relations into reversible circuit structures. This category includes phase, bit-flip, arithmetic, and Boolean function oracles, each providing a mechanism for marking  quantum states in accordance with the problem semantics. 
The circuit complexity ranges from polynomial in the input size to exponential in the worst case, depending on the nature of the encoded function.  Classically, oracle construction components are loosely analogous to predicate evaluation mechanisms, callback functions, and constraint encoders used in search and decision-making systems. From an object-oriented design perspective, quantum oracles are analogous to interfaces in that they define a contract between reusable algorithmic structures and problem-specific behaviors while concealing implementation details.  
\item \textbf{Phase Estimation Components}  These components translate phase relationships into measurable outcomes by  extracting eigenvalue information from quantum operators. This is not dependent on the semantics of the problem.  By isolating eigenvalue extraction from problem encoding and higher-level algorithm orchestration, these components support the separation of the concerns. This category includes standard quantum phase estimation (QPE), iterative QPE, and controlled unitary sequences, which recur across algorithms. The circuit complexity typically scales up to  
$O(n^2)$.  Classically, phase estimation components are analogous to the eigenvalue solvers used in numerical computing.        

\item \textbf{Variational Ansatz Components } These components realize parameterized quantum circuits used in hybrid quantum–classical optimization workflows. These components define high-level optimization objectives that are independent of concrete circuit structures. The parameters were iteratively tuned using classical optimizers. They recur across the variational quantum algorithms. They  encapsulate repeatable circuit templates with gate complexities that typically range from $O(n)$ to $O(n^2)$. By isolating the circuit structure from the optimization strategy and cost function evaluation, variational ansatz components promote the separation of concerns within the variational algorithm family.   
\end{enumerate}
The primitives are labelled according to the \emph{dominant architectural responsibility}.  This is not determined by secondary effects or internal subcomponents. Furthermore, the primary role of a circuit is characterised by parameters, such as
initialization during the lifecycle (Init), structural entanglement (Ent), semantic encoding (Sem), interference-based reshaping (Ampl), representation transformation(Basis),  eigenvalue extraction (Phase), or hybrid optimization (Param).  The decision tree shown in Figure \ref{class} ensures that each primitive satisfies exactly one criterion.  This rule resolves the apparent overlaps arising from compositional reuse. For example, circuits such as GHZ or cluster constructions may generate entanglement; however, when used to establish algorithmic preconditions, they are classified as \emph{ state-preparation }. 
The ordered decision rules guarantee that the classification is mutually exclusive and collectively exhaustive.
%Similarly, although phase estimation incorporates basis transformation internally, it is categorized as a higher-level composite module due to its eigenvalue extraction role. By grounding the labeling in lifecycle position and architectural responsibility, the taxonomy preserves conceptual clarity while acknowledging structural interdependence among primitives.

\subsection{nonfuncttional Dimensions for Classification}
\begin{figure}[t]
\centering
\includegraphics[width=0.85\textwidth]{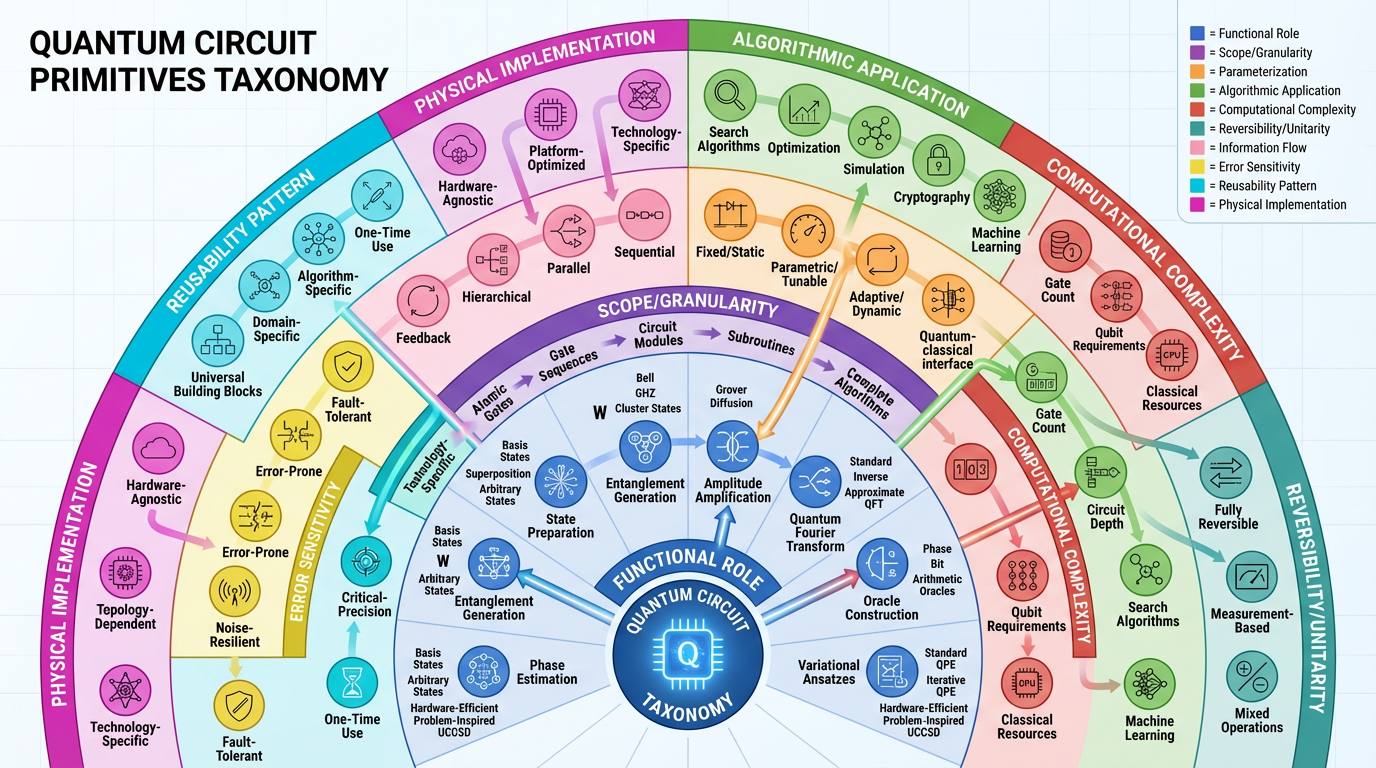}
\caption{multidimensional classification framework}
\label{fig:multidimensional}
\end{figure}

\begin{table}[t]
    \centering
    \caption{nonfuncttional dimensions for Classifications}
    \label{tab:placeholder}
    \begin{tabular}{p{0.5in}p{2in}p{2in}}
    \toprule
    Sr. No. &  Classification Parameters & Values \\ \toprule
    1 &  Scope and granularity &  Atomic Primitives, Algorithm \\  
    2 &  Degree of parameterization & fixed, structural, variational, or problem-dependent \\  
    3 &  Agorithmic application scope & universal, search, periodicity, variational, or simulation families \\  
    4. &  Computational complexity  & Often expressed in terms of gate count, circuit depth, qubit usage, and classical preprocessing cost.
  \\ 
 5 & Reversibility and unitarity properties & Yes/No \\ 
 6 &  Information-flow patterns & Local, global, hierarchical, or feed-forward \\ 
 7 &  Error Sensitivity &  Suitability for NISQ-era execution (Yes/No) \\ 
 8 &  Reusability Pattern &  Direct reuse, parametric reuse, contextual adaptation, and hierarchical composition\\ 
 9 & Physical Implementation &  Hardware Agnostic, Technologic Specific \\ \bottomrule
    \end{tabular}
    
\end{table}
In addition to their functional role, reusable quantum circuit components are characterized using a multidimensional architectural classification framework that captures orthogonal design concerns relevant to architectural decision-making ( Figure \ref{fig:multidimensional}). The framework spans nine nonfunctional dimensions, as tabulated along with their values  in Table \ref{tab:placeholder}.   Some of these dimensions capture dynamic architectural concerns, including information flow patterns,   error sensitivity, suitability for noisy intermediate-scale quantum (NISQ) era execution, and reusability patterns.   Together, these orthogonal dimensions enable architects to reason about quantum circuit multilevels from multiple perspectives to support informed trade-offs among reuse, performance, robustness, and implementability across diverse quantum software architectures.

\begin{figure}[t]
\centering
\includegraphics[width=0.85\textwidth]{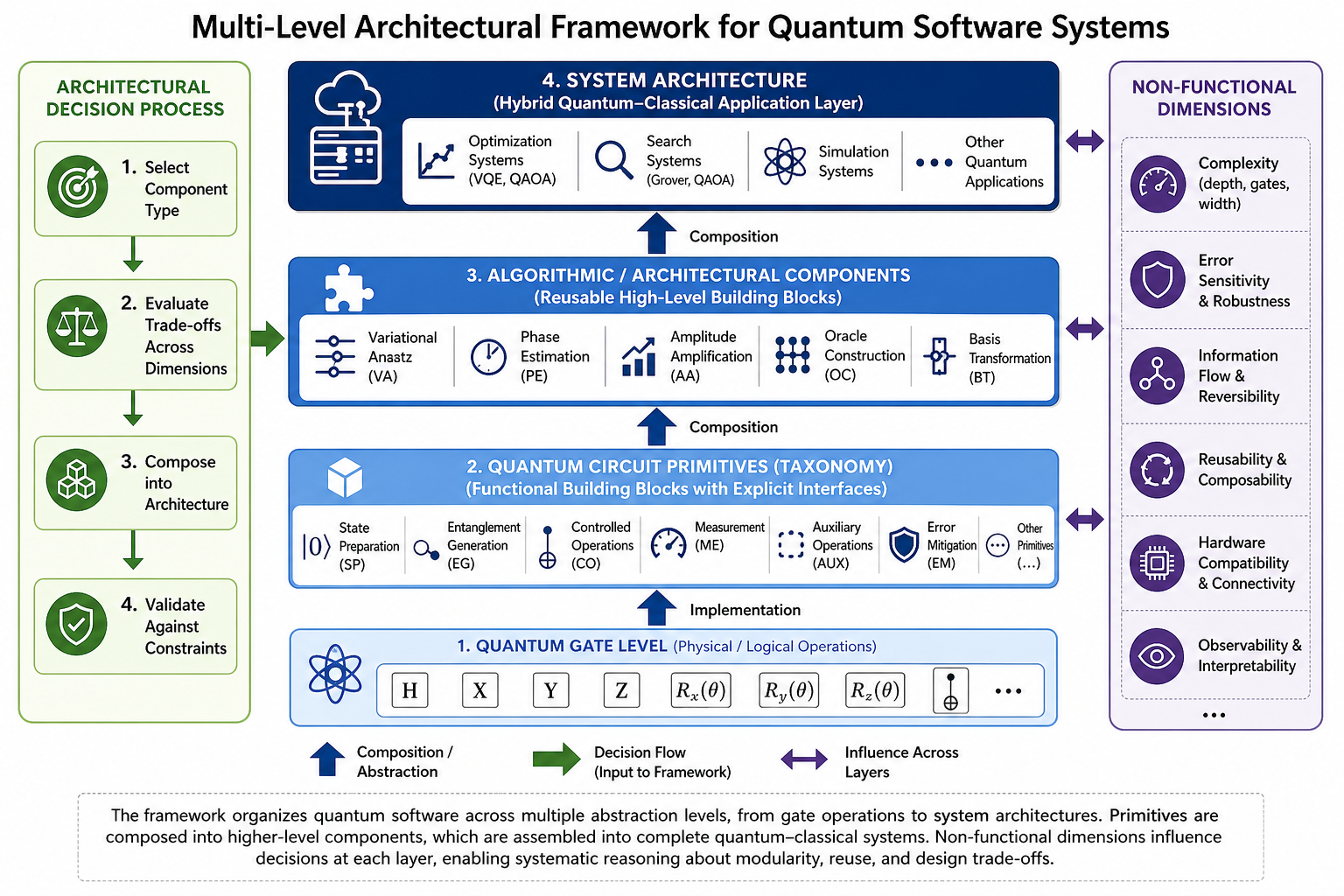}
\caption{multilevel Quantum Software Architectural  framework  (QSAF) for Quantum Software Systems}
\label{framework}
\end{figure}
 
To illustrate the decision-support capability of the proposed taxonomy, consider the architectural design of a hybrid quantum–classical optimization system that targets near-term NISQ hardware.
An architect must select a suitable variational strategy to solve a chemistry-inspired optimization problem. Two alternatives are considered:  \textit{Option A:} A hardware-efficient ansatz with shallow depth and generic entanglement patterns. \textit{Option B:} A problem-inspired ansatz (e.g., UCCSD-style structure) with a domain-specific operator encoding. Both options fall under the \textit{variational ansatz category.} However, nonfunctional dimensions expose critical architectural trade-offs. 
 UCCSD-based circuits scale polynomially but often require deeper circuits and more controlled operations. Hence, UCCSD-based circuits have higher computational complexity. In contrast, hardware-efficient ansatzes exhibit shallower depths and may be more error-sensitive and suitable for NISQ-era devices.  By explicitly reasoning across these dimensions, the architect can justify selecting a hardware-efficient ansatz for near-term noisy hardware while reserving UCCSD-style designs for fault-tolerant environments.

Figure \ref{framework} provides a system-level architectural view of \textit{Quantum Software Architecture Framework} (QSAF), showing how quantum primitives are composed into architectural components and integrated into hybrid quantum–classical systems, while nonfunctional dimensions guide architectural decision-making.

\begin{figure}[t]
\centering
\includegraphics[width=\textwidth]{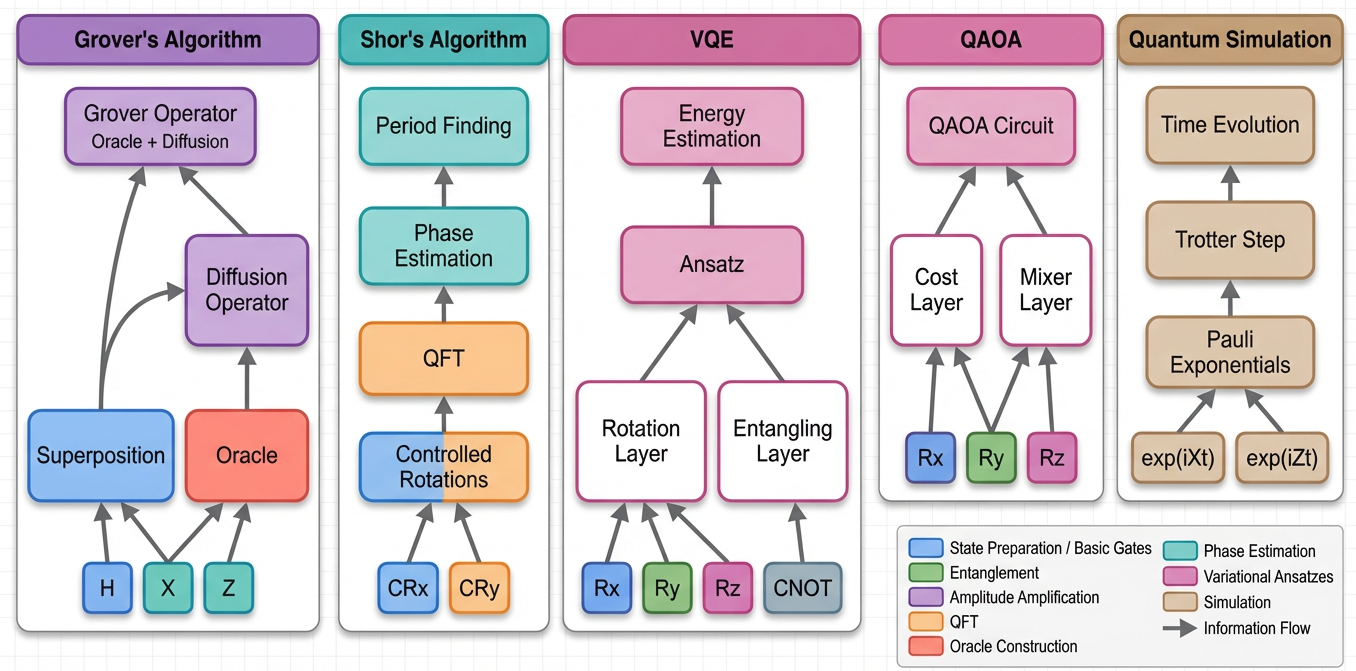}
\caption{Quantum components  across five abstraction levels}
\label{fig:hierarchical}
\end{figure}

\section{Architectural Abstraction Levels}
\label{sec:patterns}
An analysis of component usage across representative quantum algorithms reveals  regularities in both reuse and composition. As illustrated in Table~\ref{tab:heatmap}, the reusable quantum circuit components fall into three tiers of reusability. (i) A small set of universal components, including Hadamard, CNOT, single-qubit rotations, measurement, controlled operations, and SWAP, that appear across all analyzed algorithms. These components form an atomic-architectural vocabulary.  (ii) A larger group of cross-algorithm components, such as basis transformation, phase estimation, amplitude amplification, superposition preparation, and entanglement generators, are reused across multiple algorithm families. (iii)  Finally, algorithm-specific components, including Grover’s diffusion operator, modular exponentiation, UCCSD ansatz, and QAOA mixing Hamiltonians, are specialized   to individual problem classes.

Quantum software systems exhibit a consistent hierarchical composition structure spanning five abstraction levels (see Figure~\ref{fig:hierarchical}). Atomic gates form a hardware abstraction layer. They comprise  elementary primitives, such as Bell-state generators and simple superpositions, which combine to form composite primitives, such as diffusion operators or   QFT blocks. These mid-level modules are assembled into block-like algorithmic structures. For example,   Grover operators, full QFTs, or phase estimation circuits. Furthermore, these blocks comprise  complete algorithms orchestrated with classical control. This hierarchy enables  architects to reason at different abstraction levels, depending on the design intent.

\section{Discussion}
This study focuses on elevating recurring quantum circuits to first-class architectural elements in quantum software systems.  It does not claim the discovery of new quantum algorithms or primitives. Rather, its contribution lies in the architectural reframing of established circuit constructs as modular components with explicit interface definitions, abstraction boundaries, and quality-relevant properties of the circuit. Furthermore, this study does not cover distributed quantum systems and other software engineering concerns, such as runtime orchestration and deployment architecture.  The main highlights of this study are discussed below. \\  
(i) \textit{From Algorithm Classification to Architectural Concerns} Prior quantum computing literature \cite{martyn2021grand, arnault2024typology} typically classifies algorithms by their computational objectives  or mathematical structures. Such classifications explain what problems are solved, but not how reusable structures function architecturally across systems.  The taxonomy proposed here classifies circuit constructs based on architectural role, reuse tier, and nonfunctional characteristics. This shift separates the computational intent from architectural concerns.  \\ 
(ii) \textit{Interface Abstraction} Reasoning
The generalized module schema defined in Section 5
It formalizes the quantum circuit components as interface-bound modules. Prior work does not typically model them through explicit architectural contracts characterized by input/output structures, ancilla allocation, parameterization type, and information-flow patterns.
This abstraction supports architectural reasoning beyond the gate-level implementation. \\ 
(iii) \textit{Architectural Decision Support}
The value of the framework lies in enabling explicit trade-off analysis across nonfuncttional dimensions, such as complexity, error sensitivity, reuse patterns, and information flow. For example, alternative variational ansätze may be functionally equivalent, yet differ significantly in terms of circuit depth, hardware suitability, and reuse characteristics. By making such dimensions explicit, the taxonomy supports intentional design choices, rather than ad hoc circuit assembly.\\ 

Although the circuit constructs discussed in this study are   well known in the quantum computing literature, their treatment as architectural components represents a different level of abstraction. Prior work in Section 2 treats these circuits as subroutines. In contrast, this study reframes these as reusable architectural modules characterized by explicit interfaces and quality-related attributes. This shift from algorithmic description to architectural abstraction enables architecture-level reasoning to be performed. Consequently, the contribution of this study does not lie in introducing new quantum circuits but in establishing an architectural vocabulary and classification framework that supports the intentional design of quantum software systems.

\section{Conclusion}
Quantum computing introduces distinctive constraints, such as entanglement, reversibility, and the no-cloning principle, which fundamentally shape the structure of software systems. As quantum software engineering evolves beyond algorithm demonstrations towards system-level design, intentional architectural abstraction becomes increasingly important. This study reframes recurring quantum circuit structures as first-class architectural components characterized by explicit interface definitions, compositional roles, reuse tiers, and nonfunctional properties.

This study identifies reusable primitives across representative algorithm families and organizes them into a multidimensional architectural framework to establish a structured vocabulary for reasoning about modularity, composition, and trade-offs in gate-based quantum systems. The proposed module schema, functional categorization, and abstraction hierarchy collectively provide a foundation for systematic architectural design rather than ad hoc circuit assembly.

Given the early stage maturity of quantum software engineering as a discipline, architectural abstraction and conceptual structuring must precede large-scale empirical evaluations. The framework $QSAF$ proposed in this study provides a structured foundation for systematic, empirical, and tool-based validations. 
The QSAF enables architects to systematically compare alternative quantum circuit designs based on nonfuncttional trade-offs, thereby supporting informed architectural decision-making in practical quantum software systems

As quantum technologies progress towards scalable and production-oriented deployments, such architectural structuring will be essential for enabling reuse, maintainability, and long-term system evolution.
We expect the QSAF to serve as a reference model for future research on quantum software architecture and design methodologies. Future work will focus on empirical validation through case studies, tool support integration into quantum software development environments.
By shifting the focus from circuit construction to architectural design, QSAF provides a foundation for engineering scalable, reusable, and systematically designed quantum software systems
\bibliographystyle{plain}
%\bibliography{ecsabib}

\end{document}